\documentclass[review,12pt]{elsarticle}

\usepackage{amssymb}
\usepackage{float}
\usepackage{amsmath}
\usepackage{booktabs}
\usepackage{algpseudocode}
\usepackage{algorithm}
\usepackage{colortbl}
\usepackage[dvipsnames]{xcolor}
\usepackage{ulem}
\definecolor{add}{HTML}{009900}
\definecolor{remove}{HTML}{990000}

\journal{Nuclear Physics B}

\begin{document}

\begin{frontmatter}



\title{Automated Resonance Identification in Nuclear Data Evaluation}


\author[inst1]{Noah A. W. Walton}
\author[inst1]{Oleksii Zivenko}
\author[inst1]{William Fritsch}
\author[inst1]{Jacob Forbes}
\author[inst1]{Amanda Lewis}
\author[inst2]{Jesse Brown}
\author[inst1]{Vlad Sobes}

\affiliation[inst1]{organization={University of Tennessee},%Department and Organization
            city={Knoxville},
            postcode={37996}, 
            state={TN},
            country={USA}}

% \author[inst2]{Author Two}
% \author[inst1,inst2]{Author Three}

\affiliation[inst2]{organization={Oak Ridge National Laboratory},%Department and Organization
        city={Oak Ridge},
        postcode={37830}, 
        state={TN},
        country={USA}}

\begin{abstract}
% A methodological paper detailing a stable approach to automated identification of experimental resonance data. Non-convex, non-linear regression
% Global and national efforts to deliver high-quality nuclear data to users have a wide range of stakeholders, including those in national security, basic science, medical fields, and more.
Global and national efforts to deliver high-quality nuclear data to users have a broad impact across applications such as national security, reactor operation, basic science, medical fields, and more.
Cross section evaluation is a large part this effort as it combines theory and experiment to produce suggested values and uncertainty for reaction probabilities.
In most isotopes, the cross section exhibits resonant behavior in what is called the resonance region of incident neutron energy.
Resonance region evaluation is a specialized type of nuclear data evaluation that can require significant, manual effort and months of time from expert scientists.
In this article, non-convex, non-linear optimization methods are combined with concepts of inferential statistics to infer a set of optimized resonance models from experimental data in an automated manner that is not dependent on prior evaluation(s).
This methodology aims to enhance the workflow of a resonance evaluator by minimizing time, effort, and prior biases while improving reproducibility and document-ability, addressing widely recognized challenges in the field.
% The methodology 

\end{abstract}


\begin{highlights}
\item Research highlight 1
\item Research highlight 2
\end{highlights}

\begin{keyword}
Resonance \sep evaluation \sep automation \sep reproducibility
\end{keyword}

\end{frontmatter}


% =============================================
%  Introduction
% =============================================

\section{Introduction}
\label{sec:introduction}

% \textbf{reference my ND2022 paper}

% \begin{itemize}
%     \item resonance cross sections are important for applications
%     \item Default method for regression is very manual, requires prior parameters 
%     \item Model cardinality is not known a-priori
%     \item Lack of reproducibility and call for improvement
%     \item Automate the regression problem 
% \end{itemize}

The accurate modelling and simulation of nuclear systems are vital components for a wide range of applications, including national/global security, carbon-free power generation, space exploration, and medical imaging/treatment.
With the recent advances in simulation power, there is an increased need for accuracy in the fundamental nuclear data that underpin these simulations \cite{NudatNeeds_Bernstein}.

The community of nuclear data scientists is responsible for providing these fundamental nuclear data. 
This is a massive endeavor that requires global collaboration between scientists, software, and experimental facilities. 
The product is an evaluated nuclear data library, such as ENDF 8.0 \cite{ENDF8} or JEFF 3.3 \cite{JEFF3p3}, that contains mean values and uncertainties, reported as covariance, for fundamental nuclear data quantities. 
This product exists at the edge of what is known about fundamental physics and seeks to represent this information in a universal format for all reactions and isotopes. 
Often, the work is segmented into different areas of expertise, handled by different scientists, software, and facilities across the globe. 
Complete documentation and consistency across this massive project are challenges, even with modern computational hardware. 
Furthermore, much of the existing knowledge, experimental data, and software were developed during the second half of the 20th century. 
As nuclear data scientists from this era leave the field, there is a need to preserve their domain knowledge and an opportunity to modernize the available methods and tools.

These challenges have been recognized by the current nuclear data community and there have been several calls for improvement. 
New paradigms within the nuclear data pipeline, enabled by modern computational capabilities, have been a recent topic of development.
These novel efforts focus on data handling/storage \cite{FUDGE}\cite{WPEC_SG50} and evaluation \cite{TALYS_Eval}\cite{NDat_Pipeline_Schanbel}. 
A major theme throughout each of these efforts is reproducibility.
The need for reproducibility is formally addressed by the Working Party on International Nuclear Data Evaluation Co-operation (WPEC), subgroup 49 – Reproducibility in Nuclear Data Evaluation \cite{WPECSG49}.

To enhance reproducibility, automation emerges as a viable solution. 
Automation refers to a standardized computational methodology or set of procedures, ideally a software package, that reduces manual input and ensures proper execution of each incremental step.
% This concept is prolific across many applications, \cite{ISA} 
Evaluators' expertise remains indispensable, however, introducing automation into the workflow will reduce the manual workload on an evaluator and make it easier to document artisanal decisions.
Of the evaluation workflow, the inference of physics model parameters from experimental data is a prime candidate for automation.
The TALYS evaluation code \cite{TALYS_Eval} is purposely structured to lend itself to an evaluation paradigm that includes automated parameter inference.
This effort focuses on high-energy reaction evaluations and does not infer resolved resonance region (RRR) parameters.
Rather, the RRR is left to analysis codes such as SAMMY \cite{sammy} that remain manual and incremental in their approach to parameter inference.
This article presents a methodology for automating the inference of RRR parameters in support of broader efforts amongst the nuclear data community to improve reproducibility in evaluation.
An initial proof of concept preceding this work can be found in \cite{ND2022_Walton}, while the methodology presented here brings the concept to full-fidelity and enables its use in real evaluations.
% Do I need to also talk about syndat framework? 

% =============================================
%  Background section
% =============================================

\section{Background}
% \begin{itemize}
%     \item Default resonance analysis codes SAMMY and REFIT
%     \begin{itemize}
%         \item Bayesian update of prior information
%         \item Solve maximum likelihood from a weighted initial guess with assumed normal distributions
%         \item Solution scheme linearizes model by steping in the direction of the first order derivative
%         \item internal iterations for non-linearities
%         \item Spin groups are selected manually, some past codes help to loop through all spin groups but nothing has stuck
%     \end{itemize}
%     \item Recent efforts
%     \begin{itemize}
%         \item Jesse's work with MCMC
%         \item Georg's work
%         \item previous conference article with BARON optimization
%     \end{itemize}
%     \item Levenberg-Marquardt algorithm
%     \item AIC and LRT 
% \end{itemize}

\subsection{Resonance Analysis}
\label{sec:resonance_analysis}

Currently, the default resonance-analysis procedure involves using a resonance modeling and analysis code, such as SAMMY \cite{sammy} or REFIT \cite{REFIT}. 
These codes perform three primary tasks: 
1) calculate a theoretical cross section using an appropriate R-matrix formalism, 
2) mathematically model the experimental conditions that affect measured data, 
and 3) adjust the resonance parameters of the R-matrix formalism to provide a better fit to observations.

The theoretical cross section is determined by one of the formalisms of R-Matrix theory \cite{RMatrix_theory}, most commonly the Reich-Moore (RM) approximation \cite{ReichMooreApprox} for medium-to-heavy mass nuclides (A$>$20), and is parameterized by a set of resonance parameters. 
With a complete set of resonance parameters, the RM formalism can describe all energetically possible reaction cross sections for a specific incident/target particle-pair. 
Different experimental observables (i.e., capture yield, transmission, fission yield) -- none of which are direct measurements of cross section -- are used to infer these parameters.

In order to compare the \textit{cross section} model to observable data, an \textit{experimental} model must be constructed to describe how the theoretical cross section generates data under the conditions of a given experiment. 
This operation is not invertible; otherwise, the observable data could be transformed and compared to the cross section model directly. 
The mathematical implementation of these experimental models (as mentioned in item 2 above) for a range of facilities makes resonance analysis codes like SAMMY invaluable. 
Consequently, the methodology presented later in this article is built around the SAMMY code.

Item 3 above, adjusting resonance parameters to fit observations, is often framed as a Bayesian update of prior information. 
This is because there is often prior knowledge about the model from past evaluations or \textit{a-priori} theory, and evaluators are interested in updating that knowledge. 
Additionally, the output of the Bayesian framework is a posterior probability distribution on model parameters, thus allowing for an estimate of uncertainty in the evaluated parameters.

% Assuming the mean of the posterior distributions become the reported parameter estimates, the Bayesian framework also serves as a maximum-likelihood estimator. \textbf{REF MLE}
In order to solve for the Bayesian posterior in RRR evaluations, assumptions of normality (in the data and model parameters) and a linear approximation of the model allow for a closed-form solution. 
Amongst the community of nuclear data scientists, this often referred to as Generalized Linear Least Squares (GLLS).
Taking the notation from the SAMMY manual \cite{sammy}, these equations are shown below.

In order to solve for the Bayesian posterior in RRR evaluations, assumptions of normality in the data and model parameters, along with a linear approximation of the model, allow for a closed-form solution. 
This approach is often referred to as Generalized Linear Least Squares (GLLS). 
Taking the notation from the SAMMY manual \cite{sammy}, these equations are shown below.

\begin{equation}
    \label{eq:GLLS_Cov}
    M' = (M^{-1} + G^T V^{-1} G )^{-1} 
\end{equation}
\begin{equation}\label{eq:GLLS}
    \begin{aligned}
        P' & = P + M' G^T V^{-1} (D-T) \\
        % P' & = P + \frac{G^T V^{-1} (D-T)}
        %                 {M^{-1} + G^T V^{-1} G} \\
    \end{aligned}
\end{equation}

Here, $P$ and $M$ represent the mean and covariance of the prior model parameters, and the variables primed represent the posterior. 
$T$ is the theory predicted by the model with parameters $P$, and $G$ is the Jacobian matrix describing the relationship between $T$ and $P$ via a first-order Taylor expansion. 
Lastly, $D$ and $V$ are the observed data and data covariance, respectively. 

Because of the approximate, first-order relationship between $P$ and $T$, the solution's accuracy is dependent on the linearity in the space $P'-P$ and $D \pm V$. 
In an attempt to improve accuracy, SAMMY has implemented internal iterations on the evaluation of $G$ at intermediate values between $P$ and $P'$ \cite{sammy}. 
Ultimately, if the prior model is too far from the data, then these equations will not converge on an accurate solution.

There are several shortcomings to this traditional GLLS approach when applied to resonance analysis and nuclear data evaluation in general. 
These shortcomings primarily arise from neglecting considerations and/or assumptions underlying the statistical theory on which this approach is based. 
These include linearity, normality, goodness of fit, model complexity, model or data defects, and incomplete prior information. 
Each of these factors can impact the accuracy of the Bayesian posterior model.

Furthermore, practical implementation challenges often arise. 
Linearity assumptions necessitate iterative solution schemes, which still only result in local optima heavily dependent on prior information. 
This prior-dependence can bias parameter values as well as model complexity. 
Additional resonance parameters must be manually added to the model, and it is up to the evaluator to balance model complexity and goodness of fit. 
Lastly, each resonance has a quantum spin group assignment that is a discrete parameter and must be manually adjusted. 
Consequently, these discrete parameters are often heavily biased by prior evaluations.

\subsubsection{Recent Developments in Evaluation}

Recent developments in SAMMY have applied a Markov Chain Monte Carlo (MCMC) algorithm to generate posterior samples rather than solving the closed-form, deterministic equations. 
This approach addresses linearity and normality assumptions in the mathematical solution to Bayes' theorem. 
The primary motivation for introducing this alternative solution scheme was to allow for additional model components that can be used to represent model or data defects in a reproducible and reportable manner \cite{MCMC_ImperfectDataModels}.
% This method still depends on a prior as the MCMC algorithm is not an efficient global optimization algorithm for high-dimensional problems. % I don't know if this is entirely true with additions such as simulated annealing
Separately, tools developed at Brookhaven National Laboratory have been created to automatically determine good quantum spin assignments based on the likelihood given by resonance parameter distributions \cite{ML_spingroups}\cite{ResTools_BNL}. 
These tools take evaluated spin assignments and automatically return updated suggestions for the evaluator. 

Other areas of nuclear data evaluation, beyond resonance analysis, have made efforts to automate and standardize evaluators' workflows. 
Namely, the TALYS code system \cite{TALYS_Eval} is a reaction modeling and evaluation code structured to facilitate automated parameter selection.
This structure has enabled users to employ advanced techniques, 
including Gaussian processes \cite{Schnabel2018Analysis_InconsistentNuclearData}, 
Levenberg-Marquardt non-linear optimization \cite{NonLin_Fitting_NuclearReactionProducts}, 
and Monte Carlo methods \cite{MCMC_talys}\cite{NDat_Pipeline_Schanbel}.

Automation is not the only goal of the evaluation paradigm put forth by the TALYS developers; they also work towards consistency and completeness across the evaluation and validation of all isotopes. 
This is a significant effort, with reproducibility as the priority and automated parameter inference as only one aspect. 
TALYS does not perform parameter inference for the phenomenological R-matrix model in the resolved resonance region. 
Instead, TALYS, along with most nuclear data scientists around the world, continues to cite resonance analysis codes such as SAMMY as the state-of-the-art for resonance evaluation.

\subsection{Statistical Regression}

The GLLS solution provided by equations \ref{eq:GLLS_Cov} and \ref{eq:GLLS} is closely related to several other statistical regression techniques. 
This is particularly the case when the parameter estimate alone is of interest (i.e., $M'$ does not need to be interpreted). 
While the posterior covariance given by the Bayesian framework in section \ref{sec:resonance_analysis} is an important aspect of evaluation, it can be estimated in a secondary analysis that is only loosely coupled to the initial parameter estimation. 
Without the immediate need to interpret a posterior parameter covariance, the relationship between equation \ref{eq:GLLS} and other optimization techniques can be exploited. 
This section will explore these relationships using the notation introduced in the previous section \cite{sammy}.

Consider a truly linear model with a noise term given by $\epsilon$. 
If the noise around the model is normally distributed with finite variance, then the maximum likelihood estimate (MLE) of the parameters $P$ is given by minimizing the squared Mahalanobis distance between the model and observations $D$. 
The squared Mahalanobis length will be distributed as a $\chi^2$ distribution with degrees of freedom equal to the dimension of the multivariate normal distribution. 
This squared distance is commonly referred to as the $\chi^2$ statistic or goodness of fit and will be throughout this article.

\begin{equation}
    \begin{aligned}
        T       & = GP_{true} + \epsilon \\
        \epsilon& \sim N(0, V)
    \end{aligned}
\end{equation}

\begin{equation}
\label{eq:chi2}
\begin{aligned}
    P_{MLE} & = \textrm{argmin}\ (D-T)^T V^{-1} (D-T)\\
    \chi^2 & = (D-T)^T V^{-1} (D-T)    \\
\end{aligned}
\end{equation}

For the truly linear model, this optimization problem falls into the class of quadratic programming and has a closed-form solution (\ref{eq:GLS}) often called generalized least squares (GLS). 

\begin{equation}
    \label{eq:GLS}
    P_{est} = (G^T V^{-1} G)^{-1} G^T V^{-1} (D-T)
\end{equation}

In the case that the model $T$ is not linear with respect to the parameters $P$, the Jacobian matrix $G$ becomes dependent on $P$, and a closed-form solution is lost. 
However, the local behavior of the non-linear response function $T(P)$ can be approximated via a linear model (first-order Taylor approximation) in some range of finite perturbation $\Delta P$. 
This allows computable solutions to be obtained within finite perturbations that are carried out iteratively. 
This leads to the Gauss-Newton (GN) method shown in equation \ref{eq:GN}. 
The following derivations in this section are based on reference \cite{LM_algorithm}.
% This local linearization is considered a generalized linear model as it allows the non-linear response function of interest to be related to a linear model via a linking function.
% One proposed solution scheme here is iteratively re-weighted least squares 

\begin{equation}
    \label{eq:GN}
    P_{est}^{i+1} = P_{est}^i + (G^T V^{-1} G)^{-1}\ G^T V^{-1} (D-T)
\end{equation}
The functional form here is identical to the GLLS equations implemented in resonance analysis codes (\ref{eq:GLLS}), however, the interpretation is different.
The term $G^T V^{-1} (D-T)$ is inversely proportional to the derivative of the objective function (\ref{eq:chi2}).
\begin{equation}
    \begin{aligned}
        \frac{dF_{obj}}{dP} & = 2(D-T)^T V^{-1} \frac{d}{dP}(D-T) \\
                            & = -2(D-T)^T V^{-1} \frac{dT}{dP} \\
                            & = -2(D-T)^T V^{-1} G
    \end{aligned}
\end{equation}

The other term in the GN algorithm, $(G^T V^{-1} G)^{-1}$, approximates the second order derivative and rotates the direction of the step from the first order gradient.
Generally, this rotation introduced by the GN method improves the speed of convergence as compared to traditional gradient descent methods.
% Another potential solution to minimizing non-linear objective functions is gradient descent.
% For this example, one gradient descent step controlled by $\alpha$ would look like,
% \begin{equation}
%     P_{est}^{i+1} = P_{est}^{i} + \alpha (D-T)^T V^{-1} G
% \end{equation}

The Levenberg-Marquardt (LM) algorithm combines the GN algorithm with gradient descent by introducing a dampening parameter, $\lambda$, that dynamically varies the step size and the influence of the rotation term.

\begin{equation}
    \label{eq:LM}
    P_{est}^{i+1} = P_{est}^i + (G^T V^{-1} G + I\lambda)^{-1}\ G^T V^{-1} (D-T)
\end{equation}

If the $\lambda$ is small, the step resembles a GN iteration.
Conversely, if $\lambda$ is large, the step resembles gradient descent.
The dampening parameter starts large and is decreased as the objective function improves; gradient descent steps continuously move towards GN iterations as a local minima is approached.
% \textbf{Talk about additional variations on numerical implementation.}

\subsection{Variable \& Model Selection}
\label{sec:model_selection}

When inferring a model from empirical data, a common challenge is determining which explanatory variables are necessary. 
Having too many or too few explanatory variables will impact bias and variance in the inference and often decrease predictive performance \cite{Variable_Selection_Review}. 
Furthermore, in natural science applications, explanatory variables often have a physical meaning, and the goal of inference is to best describe the true underlying process. 
Appropriate variable selection in this case is vital for the interpretability of results and the subsequent physical conclusions.
% Standard practice, particularly in the natural sciences, is to tend towards simpler models \cite{Variable_Selection_Review}. 
% This stems from a more philosophical view nicely summarized by Occam's Razor .

One straightforward approach to variable selection is subset regression. 
This approach tries all possible combinations of models (subsets) from the complete set of explanatory variables ($p$). 
For each level containing $k = 1, 2, \ldots, p$ explanatory variables, the best performing model, determined by a measure of fit such as $\chi^2$, is identified. 
Then, of the set of optimal models for each level of complexity, a final model complexity can be selected using criteria such as Akaike information criteria \cite{Akaike1998}. 
While this approach is exhaustive (guaranteed to find the optimal model), it can become infeasible for large models as the search space scales as $2^p$.

Stepwise selection is an approach that significantly improves computational speed by searching down a specific trajectory. 
In forward stepwise selection, all possible $k = 1$ models are fit, and the explanatory variable that gives the most improvement determines the trajectory, i.e., the search space for the $k = 2$ models is restricted to those that contain the explanatory variable determined by the previous step. 
Backwards stepwise selection is similar; however, it starts at $k = p$ explanatory variables and iteratively removes the least impactful variable. 
In both cases, the search space scales by $1 + p(p+1)/2$. 
While these approaches are not exhaustive - there is no guarantee that the best model for each $k$ level of complexity is found - the guided search is much more computationally feasible for large models and performs well in practice \cite{ISL}.

In the variable selection approaches above, the goal is to find the best-fitting model for each level of complexity. 
The next step, selecting model complexity, is not as straightforward because the $\chi^2$ measure of fit will monotonically decrease as model complexity increases. 
In this case, information-theoretic criteria can be used to strike an appropriate balance between the goodness of fit and model complexity for the observed data. 
Some popular criteria for maximum likelihood or least squares regression problems are the Akaike Information Criterion (AIC) \cite{Akaike1998} \cite{AIC_1} and Bayesian Information Criterion (BIC) \cite{BIC_1}.
% AIC and BIC are connected to the Kullback–Leibler (KL) distance – a measure of the distance between models - a fundamental quantity in information theory and the logical basis for model selection in conjunction with likelihood inference. 
% AIC and BIC can provide a more balanced view of model selection – especially for very different complexity of the models and various dataset sizes. 

Both AIC and BIC aim to approximate the "true" model with minimal information loss as conceptualized by the Kullback–Leibler (KL) distance - a measure of the distance between models - with key differences discussed in \cite{akaike1983information}\cite{burnham2002model}\cite{8498082}\cite{1311138}. 
The key difference between the two criteria is in their purpose. 
AIC aims to minimize prediction error, focusing on the accuracy of future predictions. 
In contrast, BIC seeks to identify the best model for explaining the underlying process that generated the data \cite{burnham_2004}. 
BIC, which is based on Bayesian probability and inference, generally places a greater penalty on model complexity, favoring simpler models, particularly when the sample size is large. 
The design of BIC is such that if a true model exists within a set of candidate models, the probability of BIC selecting the correct model increases as the sample size grows. 
This feature makes BIC more preferred for large datasets, as it is more likely to avoid overfitting and choose models that are simpler and more generalizable.

AIC and BIC are calculated using equations \ref{eq:AIC} and \ref{eq:BIC} respectively, where $k$ is the number of model parameters, $n$ is the number of observations, and $\widehat{L}$ is the maximized likelihood value for the model.

\begin{equation}
\label{eq:AIC}
    AIC = 2k-2\ln(\widehat{L})
\end{equation}
\begin{equation}
\label{eq:BIC}
    BIC = k\ln(n) - 2\ln(\widehat{L})
\end{equation}

An additional term is added to the AIC criteria to adjust for small sample sizes giving the corrected AIC, or $AIC_c$.
\begin{equation}
\label{eq:AICc}
    AIC = 2k-2\ln(\widehat{L}) + \frac{2k^2 + 2k}{n-k-1}
\end{equation}

As can be seen from these equations, the second term for BIC and AIC tends to decrease as more parameters are added to the approximating model, while the first term gets larger as more parameters are added to the approximating model. 
This is the tradeoff between under-fitting and over-fitting that is quantified by these criteria. 
The model with the smallest value is the selected model, often AIC and BIC values will be reported in difference to minimum value, translating the selected model's criteria to a value of 0.0.

% =============================================
%  Methodology section
% =============================================

\newpage
\section{Methodology}

\subsection{Scope of Automation}

The target for automation in this methodology is the inferential regression procedure in resonance analysis, which identifies and characterizes resonance structures in experimental data. 
Ultimately, this is a non-convex, non-linear, inferential regression problem. As discussed in section \ref{sec:resonance_analysis}, the experimental data used in an evaluation is very heterogeneous. 
The broad range of response variables, data quality, and formatting require significant effort towards data curation. 
The expert judgment of an evaluator is invaluable during this stage; journal articles detailing the experimental conditions must be translated into SAMMY model input, discrepancies or systematically biased data must be handled, etc. 
This first stage of an evaluation is beyond the scope of this automated methodology.
The methodology presented here is meant only to augment the evaluator by taking in a curated collection of data and outputting reliable and consistent model estimates without dependence on a prior evaluation.

% Mention spin group analysis doesn't consider extra data or parameter distributions.

\subsection{Objective Function Behavior}
\label{sec:obj_func_behavior}
\begin{figure}[H]
    \centering
    \includegraphics[scale=0.2]{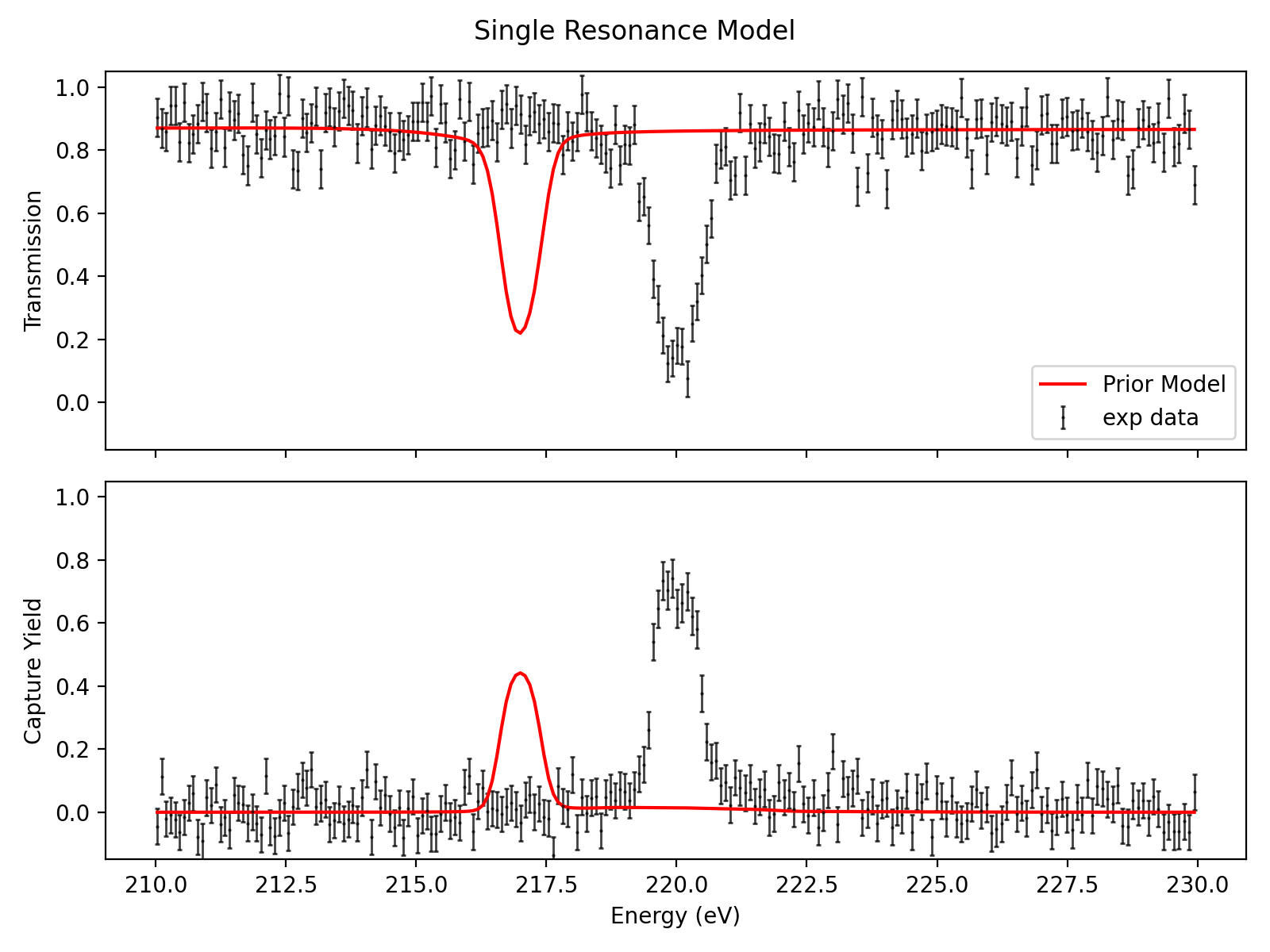}
    \caption{Fake model/observational data for mapping the objective function.}
    \label{fig:single_resonance_data}
\end{figure}

\begin{figure}[H]
    \centering
    \includegraphics[scale=0.18]{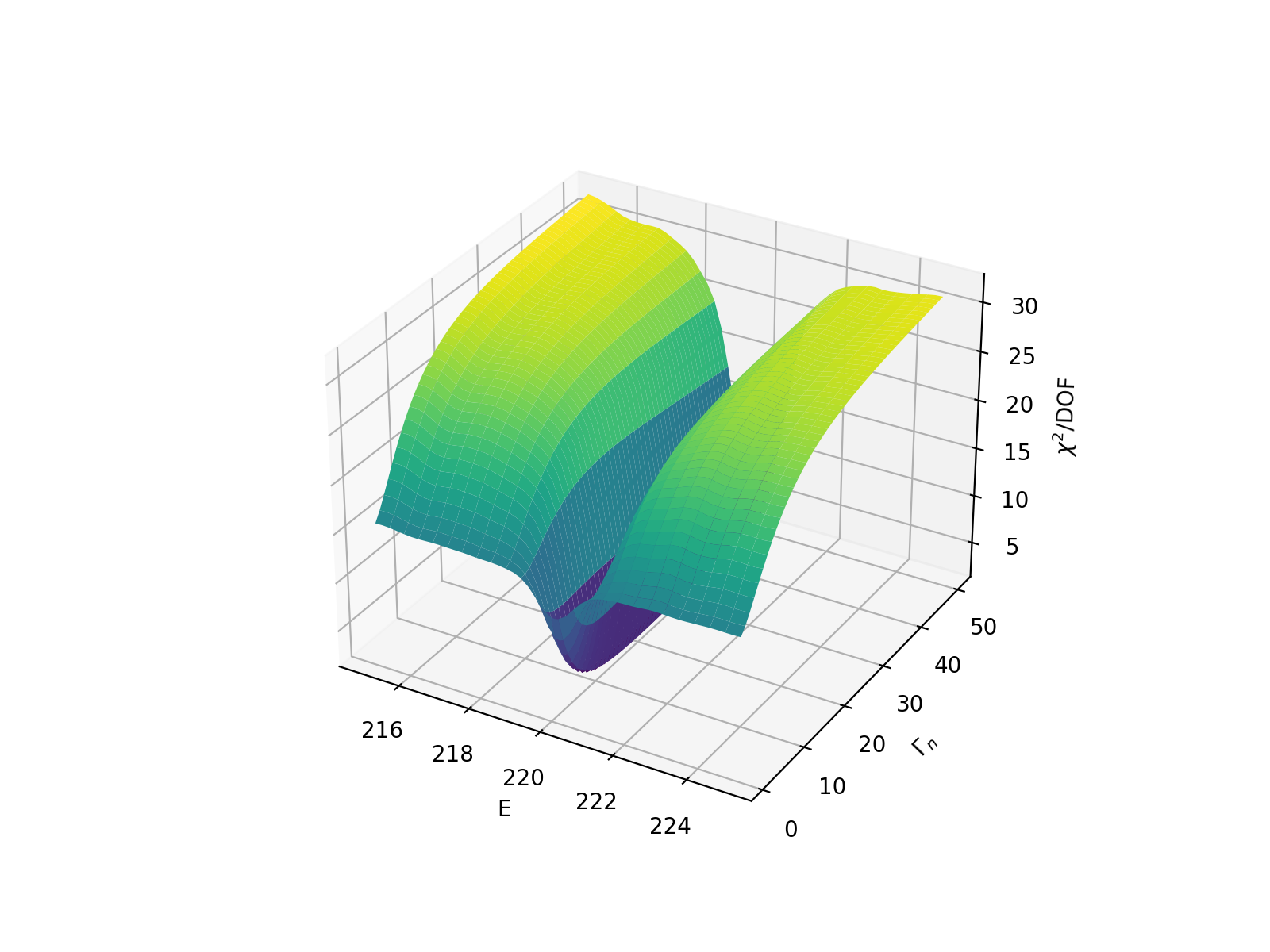}
    \caption{$\chi^2$ objective function surface over resonance energy and neutron width parameter.}
    \label{fig:objective_function_3d}
\end{figure}

\begin{figure}[H]
    \centering
    \includegraphics[scale=0.2]{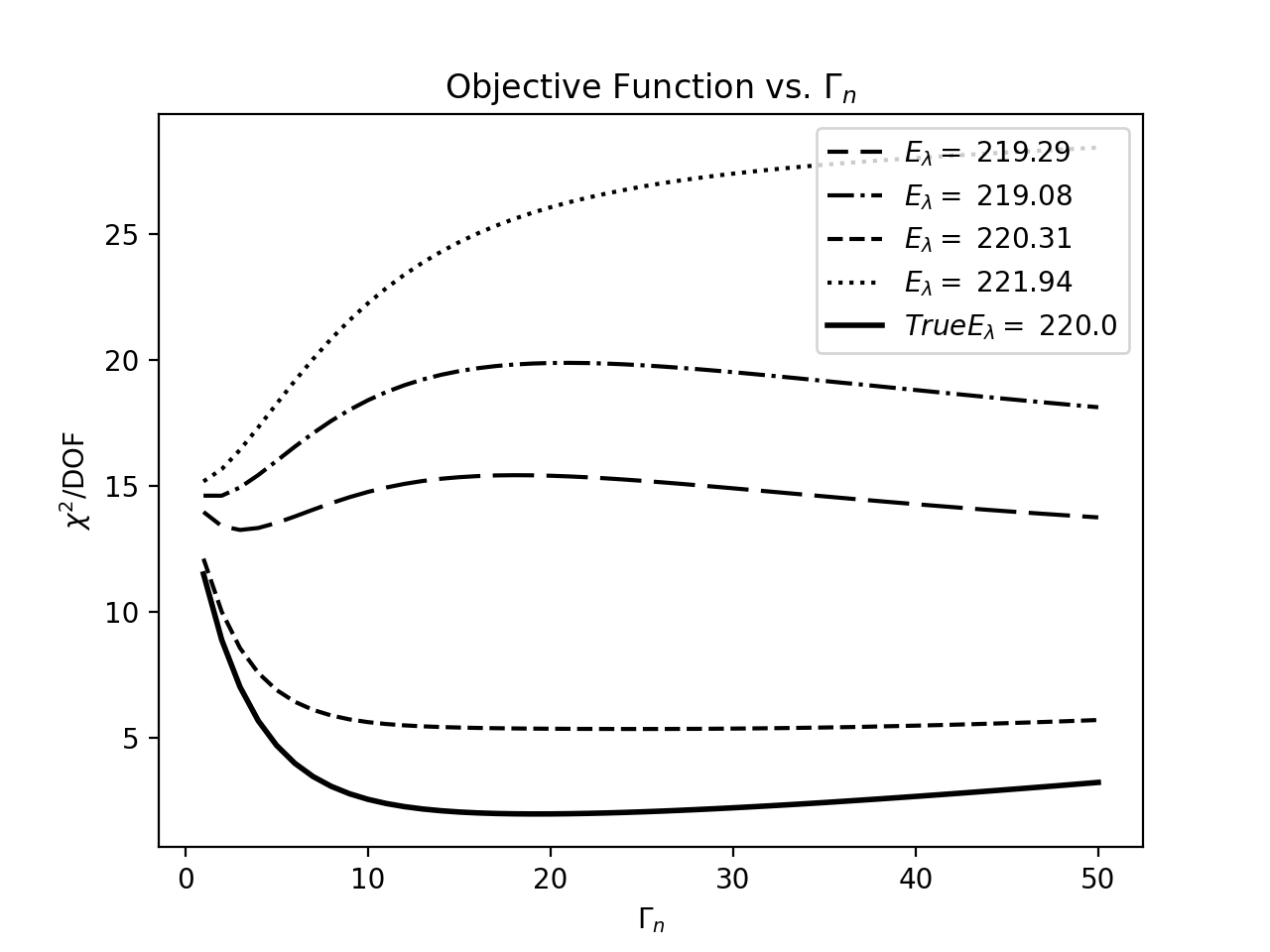}
    \caption{$\chi^2$ objective function value over neutron width parameters.}
    \label{fig:obj_vs_gn}
\end{figure}

\begin{figure}[H]
    \centering
    \includegraphics[scale=0.2]{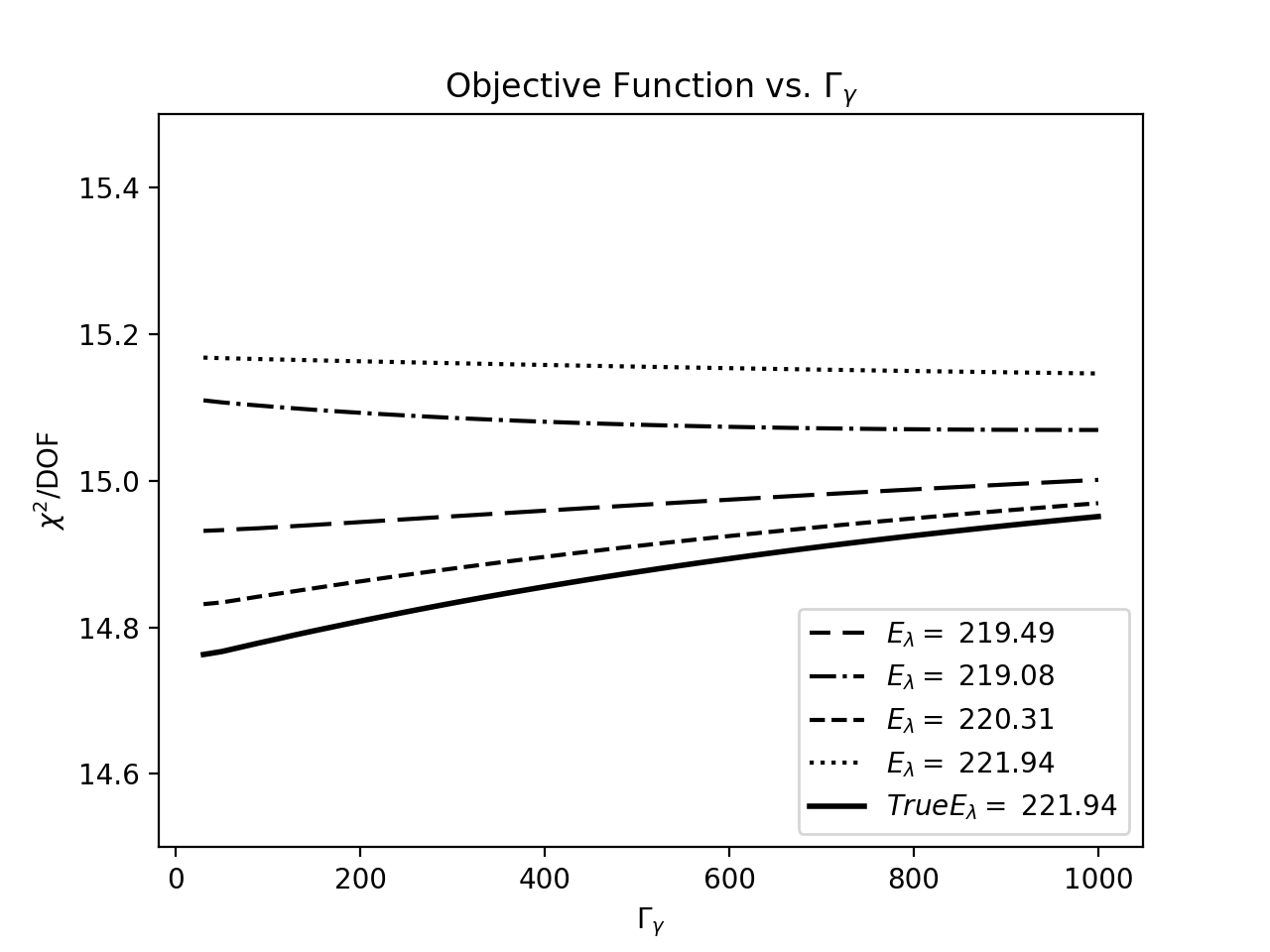}
    \caption{$\chi^2$ objective function value over capture width parameter.}
    \label{fig:obj_vs_gg}
\end{figure}

The inferential regression that is the goal of this automated methodology can be framed as an optimization problem with the objective of minimizing the squared Mahalanobis distance ($\chi^2$ statistic) for a given model.
This section gives a simple example to illustrate the behavior of this objective function with respect to the model parameters.
Figure \ref{fig:single_resonance_data} shows some observational data and a prior model, both are dummy data (with realistic values) generated for this example.
The resonance model is parameterized with energy location ($E_{\lambda}$) and width parameters ($\Gamma_{\gamma}$ and $\Gamma_n$).
% The prior model gives a starting point for a non-linear solution-iteration scheme. 
A grid search over the parameters can be done to map the behavior of the objective function. 
Figure \ref{fig:objective_function_3d} shows the objective function with respect to $E_{\lambda}$ and $\Gamma_n$ in 3D.
For any given value of width parameters, the convex region exists only narrowly around $E_{\lambda}$. 
Furthermore, as $E_{\lambda}$ moves away from the minima of the convex region, the objective function becomes non-convex with respect to the other two parameters as shown in figures \ref{fig:obj_vs_gn} and \ref{fig:obj_vs_gg}. 
% The objective function with respect to $\Gamma_n$ is mostly convex regardless of the $E_{\lambda}$, with local minima near zero if as $E_{\lambda}$ moves away from it's convex region.
While this is a simplified example, it captures the general relationship between resonance parameter and objective function.
These non-convex and non-linear relationships are the motivation for the algorithmic approach to optimization described in the following sections.

% To better visualize the gradient behavior with respect to the width parameters, figures \ref{fig:obj_vs_gn} and .

\subsection{Exploiting SAMMY}
\label{sec:SammyLM}
As aforementioned, this methodology is built around the resonance analysis code, SAMMY \cite{sammy}.
Instead of a prior parameter covariance ($M$), SAMMY allows a user-input, relative, prior parameter covariance scaled by $\zeta$.
If this option is used, SAMMY solves the following equation (with internal iterations to improve linear assumptions on G):
\begin{equation}\label{eq:GLLS_mod}
    \begin{aligned}
        P' & = P + ( \textrm{diag}(\zeta P)^{-1} + G^T V^{-1} G )^{-1}\  G^T V^{-1} (D-T) \\
        % P' & = P + ( (\mathbf{I} * \zeta P)^{-1} + G^T V^{-1} G )^{-1}\  G^T V^{-1} (D-T) \\
    \end{aligned}
\end{equation}
Each time SAMMY is called, this equation is solved. 
Replacing $\zeta$ with $\frac{1}{\lambda}$, this becomes identical to a Levenberg-Marquardt (LM) step with a dampening factor that is weighted by the magnitude of each parameter.
By iteratively calling SAMMY, capturing the output, and dynamically varying the $\zeta$ input, the LM algorithm can be implemented, providing a robust method for finding local minima of the non-linear objective function.
% Talk briefly about the other paper that uses LM \cite{NonLin_Fitting_NuclearReactionProducts}
% This algorithm is more robust at finding local minima for non-linear objective functions than iterating SAMMY with a static $\zeta$, which was shown to be equivalent to the Gauss-Newton algorithm \textbf{ref}.

% through one of three different internal solution schemes that iterate on G to help with non-linearity (see \cite{sammy}).

\subsection{The Algorithm}
The complete, automated resonance identification algorithm can be split into two major steps;
1) regression with an initial feature bank composed of an excess of resonances, 
2) backwards step-wise selection to iteratively eliminate resonances. % and get optimized parameters for each level of model complexity.
The purpose of step 1 is to address the non-convexity of the problem.
Further details are given in section \ref{sec:FB}, but the effective result is a fitted model of high-complexity.
Domain knowledge about resonance level spacing \cite{wigner_distr_original} (and the general philosophy of Occam's razor that underlies scientific inference) tells us that this result is likely over-fitting.
Thus, the second step is needed to reduce model complexity.
Model complexity has a convenient interpretation as each resonance corresponds to a discrete level in the compound nucleus, i.e., model complexity is synonymous with the number of resonances.
The backwards selection scheme seeks to find the optimal model for each integer number of resonances; starting from a many-resonance model and reducing down to zero. % (or some reasonable low number of resonances determined by the user).
% During this step, optimal spin group assignments (based on $\chi^2$ alone) are also determined.
Further details are given in section \ref{sec:elim}.

\begin{figure}[H]
    \centering
    \includegraphics[scale=0.65]{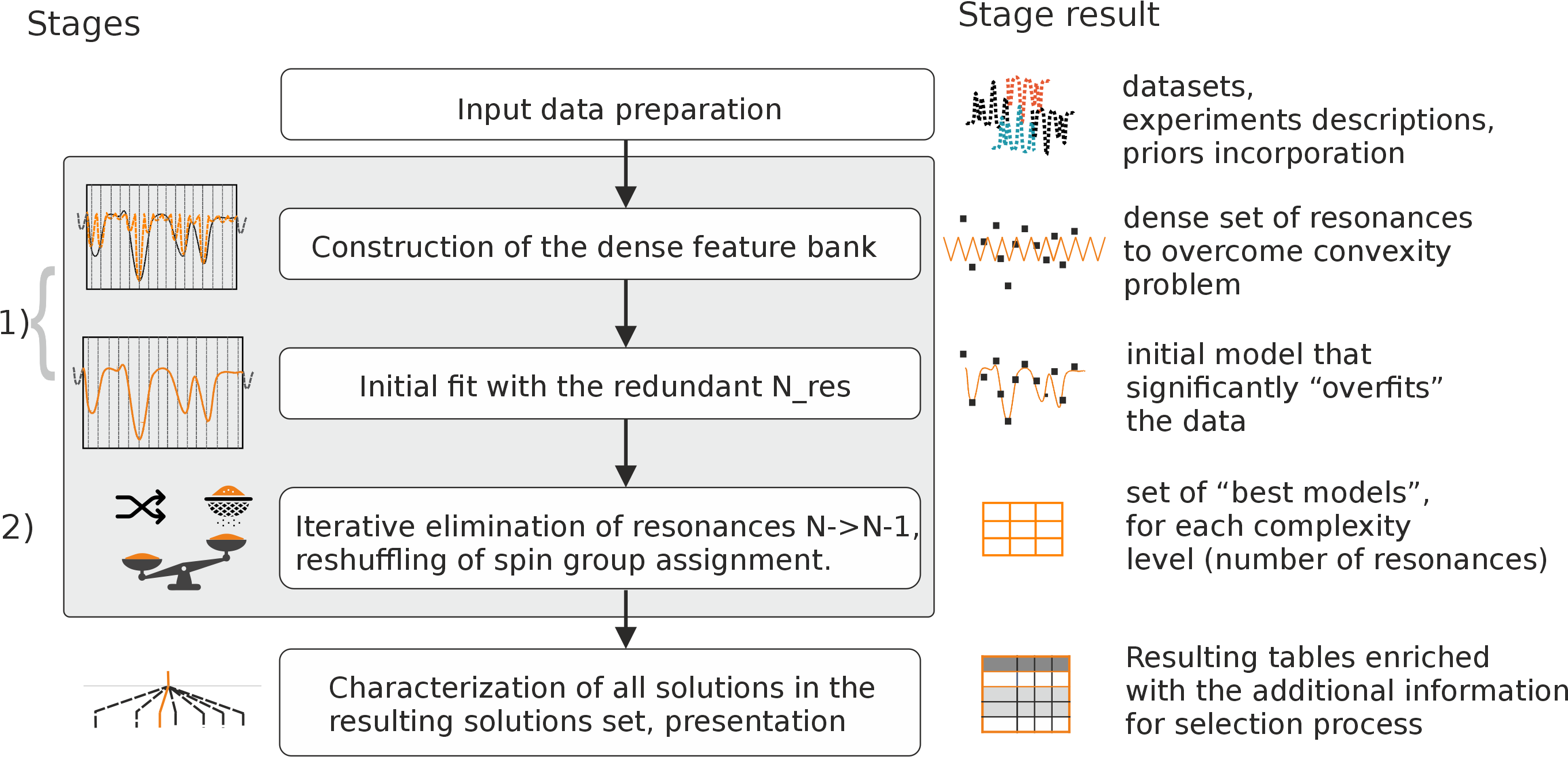}
    \caption{Automated fitting algorithm workflow diagram.}
    \label{fig:algflow}
\end{figure}

In all steps, optimization refers to the minimization of the squared Mahalanobis distance ($\chi^2$) via the LM algorithm for non-linear models.
% The authors recognize that evaluators often consider additional data (e.g., gamma multiplicity, resonance integral) or additional likelihood models (e.g., Porter Thomas \& Wigner parameter distributions). 
% These are not included in the optimization scheme presented here, but will be considered in future work.
Figure \ref{fig:algflow} shows the overall workflow with the automated fitting algorithm in the gray-shadded box.
Rather than automating the selection of model-complexity, this decision is left to the expertise of the evaluator.
The final output of this methodology is a report of the optimized model(s) for each level of model complexity (number of resonances).with total angular momentum \& parity 
% Model selection criteria (AIC, BIC, \& others) are reported for the selected model for each level of complexity. 

% The AIC and BIC criteria do not always select the same model complexity and the incorporation of parameter distributions is left for future work.
% Furthermore, it is unlikely that these decisions can be automated for all isotopes and data in a "one-size-fits-all" way.
% Therefore, this methodology is not claiming this decision, rather, it is left to the expert judgement of the evaluator who can better incorporate domain knowledge.

\subsubsection{Feature Bank Initialization}
\label{sec:FB}

Implementation of the LM algorithm gives robust solutions to local minima for the non-linear objective function, however, it does not address the non-convexity of the problem.
% This is particularly a challenge when trying to infer the energy location ($E_{\lambda}$) of a resonance.
Global minimization of non-convex problems can be a challenging problem, often requiring stochastic and/or heuristic algorithms \cite{OptforML}.
This becomes more difficult when the cardinality of the problem is not known.
To address this, domain knowledge is used to engineer a starting point that forces convexity in the objective function.
This starting point is considered a ``feature bank" where each feature corresponds to a resonance structure at some energy, $E_{\lambda}$.
% The feature bank consists of $E_{\lambda}$ and $\Gamma_g$ features as these are the primary non-convex parameters.
Of course, exact regions of convexity will depend on the specific data and model at hand.
Generally, a dense, uniform grid of resonance energy locations (e.g., $>$ 1 per eV) for each possible spin group ($J^{\pi}$) will ensure convexity with respect to $E_{\lambda}$.
However, if the average level spacing is very small a more dense grid may be necessary and spacing can be determined by some lower quantile of the respective Wigner distribution.
The initializing values for $\Gamma_n$ should generally be small to ensure they are on the left side of curves shown in figure \ref{fig:obj_vs_gn}.
This can be given by some lower quantile of the respective Porter Thomas distributions \cite{PT_original}.
The choice of initializing values for $\Gamma_\gamma$ are less straight forward; convexity in this parameter is highly sensitive to the other parameters $E_{\lambda}$ and $\Gamma_n$, while the objective function itself is not very sensitive.
Consequently, this parameter is not optimized until near-optimal values have been found for $E_{\lambda}$ and $\Gamma_n$.
Until that point, the prior estimated average width is used to populate the resonance feature bank with $\Gamma_\gamma$ values.
The algorithmic procedure is outlined in algorithm \ref{alg:initialFB}.
% This is common in many other resonance evaluation schemes

\begin{algorithm}[H]
\caption{Initial feature bank optimization}\label{alg:initialFB}

\begin{algorithmic}[0]

\State $\vec{E}_{\lambda}
        \gets 
        \{E_0, E_1, E_2, \ldots, E_n\} 
        $

\State $\vec{\Gamma}_{\gamma} 
        \gets 
        \{ \langle\Gamma_{\gamma}\rangle, \langle\Gamma_{\gamma}\rangle, \langle\Gamma_{\gamma}\rangle, 
        \ldots, \langle\Gamma_{\gamma}\rangle\}
        $

\State $\vec{\Gamma}_{n}     
        \gets  
        \{\Gamma_{n, 0}, \Gamma_{n, 1}, \Gamma_{n,2}, 
        \ldots, \Gamma_{n, n}\}
        $

\While{un-converged} 
\State $ \underset{\vec{\Gamma}_n}{\text{min}} \ 
\chi^2 $ (fit with only neutron width)
    % \If{$\Gamma_{n,i} < \theta$}
    %     \State $\vec{E}_{\lambda} \gets \vec{E}_{\lambda} \setminus E_i$
    %             (remove feature)
    %     \State $\vec{\Gamma}_{\gamma} \gets \vec{\Gamma}_{\gamma} \setminus \Gamma_{\gamma, i}$
    %             (remove feature)
    %     \State $\vec{\Gamma}_{n} \gets \vec{\Gamma}_{n} \setminus \Gamma_{n,i}$
    %             (remove feature)
    % \EndIf
\EndWhile

\While{un-converged} 
\State $ \underset{\vec{E}_{\lambda}, \vec{\Gamma}_{\gamma}, \vec{\Gamma}_n}{\text{min}} \ 
\chi^2 $ (fit with all resonance parameters)
    % \If{$\Gamma_{n,i} < \theta$}
    %     \State $\vec{E}_{\lambda} \gets \vec{E}_{\lambda} \setminus E_i$
    %             (remove feature)
    %     \State $\vec{\Gamma}_{\gamma} \gets \vec{\Gamma}_{\gamma} \setminus \Gamma_{\gamma, i}$
    %             (remove feature)
    %     \State $\vec{\Gamma}_{n} \gets \vec{\Gamma}_{n} \setminus \Gamma_{n,i}$
    %             (remove feature)
    % \EndIf
\EndWhile

% \For{$\Gamma_{n,i}$ in $\vec{\Gamma}_{n}$}
%     \If{$\Gamma_{n,i} < \theta$}
%         \State $\vec{E}_{\lambda} \gets \vec{E}_{\lambda} \setminus E_i$
%                 (remove feature)
%         \State $\vec{\Gamma}_{\gamma} \gets \vec{\Gamma}_{\gamma} \setminus \Gamma_{\gamma, i}$
%                 (remove feature)
%         \State $\vec{\Gamma}_{n} \gets \vec{\Gamma}_{n} \setminus \Gamma_{n,i}$
%                 (remove feature)
%     \EndIf
% \EndFor

\State \Return{
$\vec{E}_{\lambda}, \vec{\Gamma}_{\gamma}, \vec{\Gamma}_{n}$
optimized parameters}
\end{algorithmic}
\end{algorithm}

\noindent In the above pseudocode, minimization of $\chi^2$ refers to the the LM algorithm implemented around SAMMY and convergence is determined by some threshold of improvement in the objective function.

\subsubsection{Feature Elimination \& Model Selection}
\label{sec:elim}

The initial feature bank solve will give an overfit model with many resonances ($p$).
Starting with a full set of $p$ explanatory variables naturally lends itself to backwards step-wise selection.
Additionally, forward step-wise selection is less stable for this application because the first steps (few-resonance models) may fall outside of the region of convexity in the objective function.

% The algorithm for feature selection is this:
% \begin{algorithmic}
% \State $i \gets 10$
% \If{$i\geq 5$} 
%     \State $i \gets i-1$
% \Else
%     \If{$i\leq 3$}
%         \State $i \gets i+2$
%     \EndIf
% \EndIf 
% \end{algorithmic}

% The backwards step-wise approach is computationally greedy compared to the exhaustive best-subset selection approach, however, the exhaustive approach quickly becomes computationally infeasible as the number of resonances increases.
% Additional options for more-or-less greedy selection trajectories can be included in the implementation of this backwards selection algorithm.
% For example; 
% relaxed convergence criteria in the non-linear optimization can be used during this stage, 
% or, 
% a threshold for total resonance width can be set to automatically eliminate negligibly-small resonance features in a single step.
% These options are left to the user to decide based on domain knowledge.
% In each case, the less-greedy the approach gives more confidence that a global optima has been found.
% However, as shown later in the results, greedier approaches often select the same or similar models as the more exhaustive approaches and give solutions much quicker.

The backwards selection routine gives optimized models for each level of complexity.
As aforementioned, model selection is left to the expertise of the evaluator.
However, model selection criteria such as AIC and BIC can be used to help guide that decision.
These criteria are used with a likelihood function given only by the goodness of fit to the data as shown in equation \ref{eq:AIC_chi2}.
\begin{equation}
\label{eq:AIC_chi2}
    AIC = 2k - 2\ln(\widehat{L}) = 2k + \chi^2 
\end{equation}
This model selection criteria, however, neglects a-prior domain knowledge given by resonance parameter distributions.

\newpage
\section{Results}
% \begin{itemize}
%     \item Method works on full-fidelity experimental data, demonstrated on Jesse's Ta-181 measurements
%     \item test and show optimal speed with internal iterations, LM parameters, etc (should internal iterations be 0 if using LM?)
%     \item Able to get near identical same solution as manual evaluation starting from JEFF 3.3 prior
% \end{itemize}

The automated resonance identification methodology presented in this report was demonstrated on a set of Ta-181 measurement data taken at Rensselaer Polytechnic Institute \cite{BrownThesis}\cite{Ta181_measurements_Brown}.
These data consist of 3 transmission measurements and 2 capture measurements at different target thicknesses. 
This data is the only data that will be used in this demonstration.
To reiterate, this automated methodology does not replace the evaluators responsibility to carefully curate experimental data to be used in an evaluation.
While this demonstration uses only these most recent Ta-181 measurements, the same could be done for an evaluation that considers additional, past measurements.

In order to make a statement about the quality of this automated methodology, results are compared to a more traditional approach: optimizing parameters starting from a prior evaluation (JEFF 3.3 \cite{JEFF3p3} and ENDF 8.0 \cite{ENDF8}).
For these optimizations it is assumed that the prior covariance is large (uninformative) and that the spin group assignments do not change.
Then the LM algorithm is used to find a local minimum with the same precision (convergence criteria) as used for the final step in the automated solutions.
It should be noted that the prior evaluations did not have access to the data used here.

Lastly, this demonstration considers one ``window" of incident neutron energy of the resolved resonance region.
This allows for more digestible results, however, this methodology could be applied to the entire resonance region for Ta-181 at the cost of additional computation time.
For both the automated and manual results, external resonances at fixed energies just outside of the window were included to represent the in-window contribution from distant resonances.

\begin{figure}[H]
    \centering
    \includegraphics[scale=0.25]{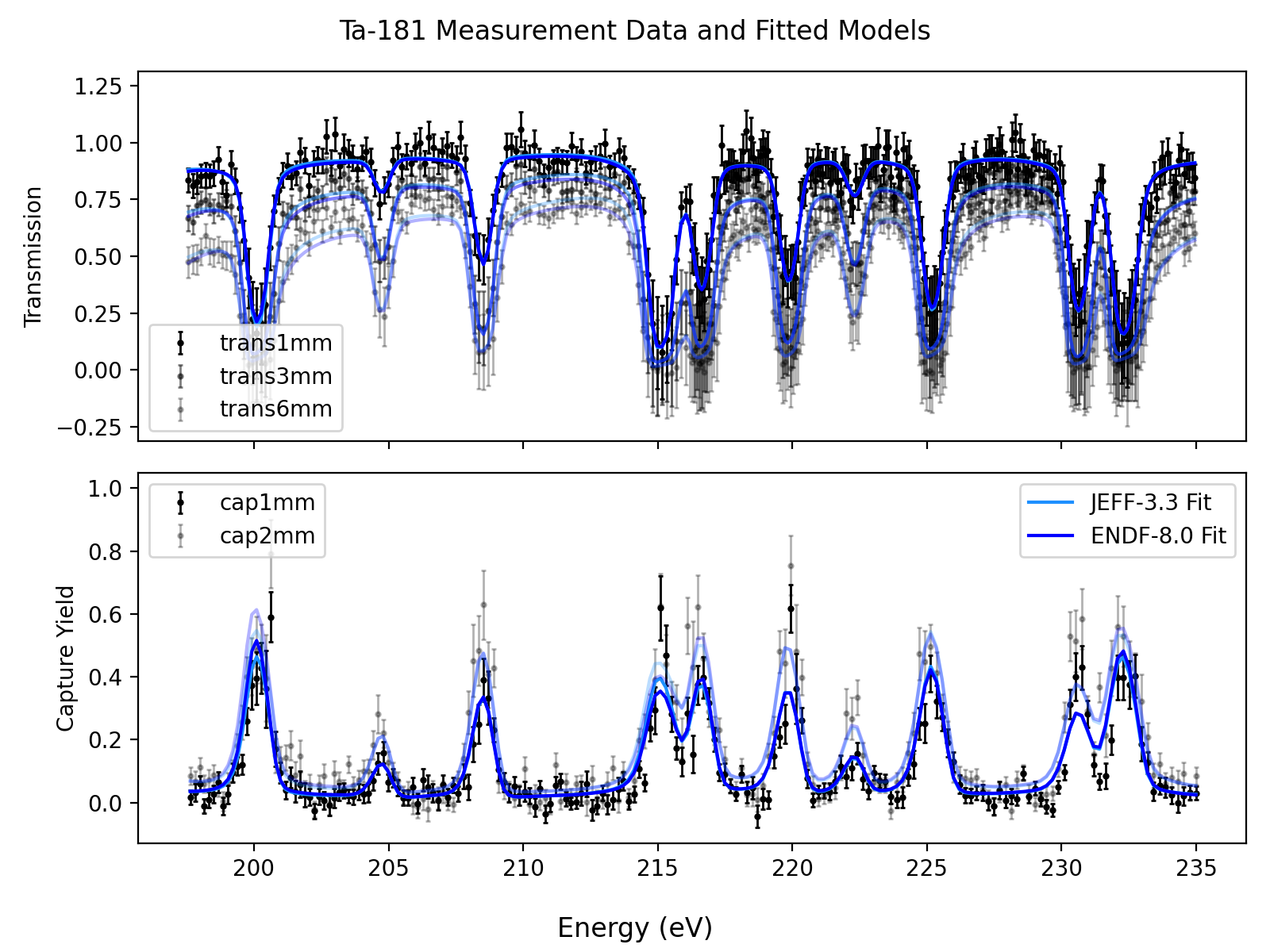}
    \caption{Window (40eV) of RRR with observational data and fitted models. 
    The fitted models start from a prior given by ENDF-8.0 or JEFF-3.3 evaluations.
    The results are almost identical, however, differences in resonance spin assignment give slightly different fits.}
    \label{fig:pw_comparison}
\end{figure}

Figure \ref{fig:pw_comparison} shows the window of experimental data and fitted models starting from prior evaluations: ENDF-8.0 and JEFF-3.3.
The energy window of interest is between 195 and 235 eV.
The prior and posterior resonance parameters are reported in the appendix.

Table 1 gives information on each candidate model given by the automated methodology described in this article.
The same information for models fitted from JEFF-3.3 or ENDF-8.0 priors are given for comparison.
As expected, the more complex models achieve a lower value of $\chi^2$.
Model selection criteria are intended to help combat this, however, for this application, the benefit had in $\chi^2$ outweighs the penalty from additional resonance parameters as the AIC and BIC criteria both point to a model that is more complex than JEFF-3.3 or ENDF-8.0.
In this case, additional, a-priori information about resonance level spacing may aid in model selection.
To represent this information, the resonance locations for each spin group were tested against the Wigner distribution using a 1-sample Kolmogorov-Smirnov test \cite{KS_test}.
Significance values are reported in columns $ks_{D1}$ and $ks_{D2}$ (for spin groups 1 and 2 respectively).
Lower values indicate that the spacing between resonances in that model is less likely. 
The traditional significance level of 0.05 would reject the hypothesis that the higher complexity models (11-13 resonances) are statistical samples from the Wigner distribution.
Note that these tests also reject the spin assignments given by the JEFF-3.3 prior.

\begin{table}[H]
\label{tabel:output}
\begin{tabular}{llrrrrrrrr}
\toprule
% \frac{\chi^2}{N_\mathrm{data}}
Prior  & \#R & $\frac{\sum\chi^2}{N_\mathrm{data}}$ & $\sum \chi^2 $ & AIC & BIC & $ks_{D1}$ & $ks_{D2}$ \\
\midrule
None     & 13.00 & 1.18 & 1531.96 & 5.86    & 21.37   & 0.17 & 0.01 \\
None     & 12.00 & 1.18 & 1532.10 & 0.00    & 0.00    & 0.25 & 0.01 \\
None     & 11.00 & 1.22 & 1579.93 & 41.83   & 26.24   & 0.01 & 0.13 \\
None     & 10.00 & 1.24 & 1605.04 & 60.94   & 29.84   & 0.13 & 0.14 \\
JEFF-3.3 & 10.00 & 1.31 & 1703.43 & 159.32  & 128.30  & 0.97 & 0.01 \\
ENDF-8.0 & 10.00 & 1.26 & 1643.58 & 99.47   & 68.44   & 0.53 & 0.20 \\
None     & 9.00  & 2.06 & 2677.26 & 1127.15 & 1080.55 & 0.28 & 0.14 \\
None     & 8.00  & 2.30 & 2992.00 & 1435.90 & 1373.79 & 0.21 & 0.41 \\
\bottomrule
\end{tabular}
\caption{Results table comparing goodness of fit measures for different models.
The rows of this table with 'None' as the prior are the output of the automated methodology detailed in this article.
The same goodness of fit measures for the models that used prior evaluations are included for comparison.}
\end{table}

Taking the same model complexity as given by the prior evaluations, the automated methodology identifies the same resonance energy levels with different spin assignments, resulting in slightly different widths and a better $\chi^2$ fit to the data for the same model complexity (10 resonances).
This is visualized in figure \ref{fig:autofit_pw} and exact resonance parameters are given in the appendix.

\begin{figure}[H]
    \centering
    \includegraphics[scale=0.25]{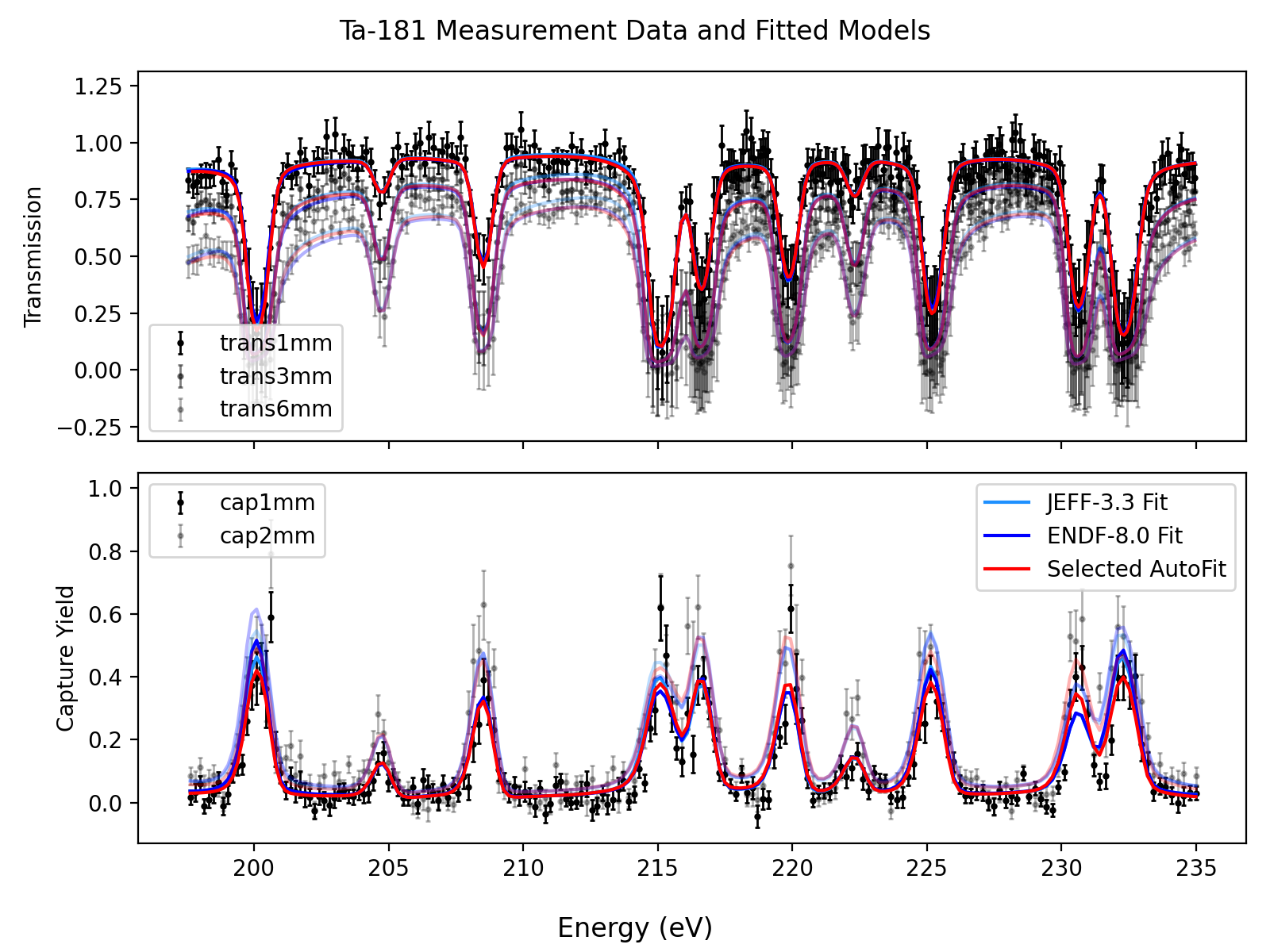}
    \caption{
    Window (40eV) of RRR with observational data and fitted models comparing different fitting methods -- starting from a chosen prior or using the automated methodology presented here. 
    The selected automated fit was chosen to be of the same model complexity as the models that started from prior evaluations (10 resonances). 
    The model given by the automated approach is nearly identical, identifying very similar resonances with differing spin assignments.
    }
    \label{fig:autofit_pw}
\end{figure}

% \section{Additional Results}

\section{Conclusions \& Future Work}

This article presented a methodology that leverages feature engineering, the LM algorithm for non-linear optimization, and backwards step-wise regression to automate the identification and characterization of resonances in experimental cross section data. 
One major strength of this methodology is that it can be implemented using SAMMY, a tool that many resonance evaluators already use.
Implementation of this methodology would reduce manual effort and remove potential biases caused by the selection of a prior as a starting point, which, can particularly affect the number of resonances and spin assignments in a resonance evaluation.

The primary limitation of this methodology arises from the backward step-wise variable selection approach. 
Primarily, this method offers no assurance of discovering the global minima for any given model complexity. 
A solution to this challenge is best-subset selection, which guarantees locating the global minima; however, its practicality diminishes rapidly with the increasing number of resonances. 
Transitioning from best subset to step-wise represents a shift toward a more greedy algorithm for variable selection. 
Even within the step-wise algorithm, there exist variations in greediness. 
Empirical observations during the development of this algorithm reveal the following insights. 
The level of greediness tends to influence the models encountered at high complexities, particularly when many of the resonance features are merely fitting noise in the data.
Nevertheless, as the selection algorithm progresses toward model complexities deemed more reasonable (as determined by AIC/BIC and a-priori knowledge about resonance level spacing), different levels of greediness converge to nearly identical models, irrespective of the path taken through high-complexity scenarios.

Additionally, as the energy window of interest expands, even the more aggressive versions of the step-wise algorithm encounter computational limitations. 
Future enhancements of this methodology could incorporate regularization or dimensionality reduction techniques to address this constraint. 
However, these approaches lie beyond the current scope of work for two main reasons: 
1) their application is not straightforward for models that exhibit non-linearity in the explanatory variables, and 
2) integrating these additions would deviate from the exploitation of SAMMY input (section \ref{sec:SammyLM}), rendering the overall methodology less accessible to current resonance evaluators.

Another directly related avenue for future research is the extension of automation to encompass the selection of model complexity. 
Presently, the evaluator must still choose a candidate model complexity as this method generates a range of potential options. 
Although this aspect is deferred to future work, the current methodology contributes to the overarching objective of enhancing reproducibility/report-ability and alleviating the manual workload on evaluators.

% =============================================
%  Appendices and bibliography section
% =============================================

\newpage
\appendix

\section{Reported Resonance Parameters}
The first table in this appendix gives parameter and spin group definitions as well as the averages used for parameter distribution tests.
The following tables show the fitted resonance parameters corresponding to results in this article. 
For the models evaluated from a prior evaluation, both prior and posterior parameters are given with the posterior parameters indicated by an apostrophe ($'$).
For the models given by the automated methodology, prior parameters are not reported.
In the tables with model parameters, those highlighted represent the external resonances appended to capture contribution from resonances outside the window.
 
% # J = 3, <D> = 9.0030 eV, <gn2> = 452.56615 meV, <gg2> = 32.0 meV
% # J = 4, <D> = 8.3031 eV, <gn2> = 332.24347 meV, <gg2> = 32.0 meV

\begin{table}[H]
\label{tabel:parameter_definition}
% \caption{Resonance parameter definition and units
% }
\begin{tabular}{ll}
\toprule
Parameter & Description \\
\midrule
$E_{\lambda}$       & Resonance energy in eV \\
$\Gamma_{\gamma}$   & Neutron width in meV evaluated at resonance energy ($E_{\lambda}$ ) \\
$\Gamma_n$          & Eliminated capture width in meV evaluated at resonance energy ($E_{\lambda}$ ) \\
$\gamma_{\gamma}^2$ & Reduced eliminated capture width in meV, ($\frac{\Gamma_{\gamma}}{2*P(E_\lambda)}$) \\
$\gamma_{n}^2$ & Reduced eliminated capture width in meV, ($\frac{\Gamma_{n}}{2*P(E_\lambda)}$) \\

$\langle D \rangle$ & Average level spacing for given spin group    \\
$\langle \gamma_{\gamma}^2 \rangle$ & Average reduced eliminated capture width for given spin group \\
$\langle \gamma_{n}^2 \rangle$ & Average reduced neutron width for given spin group \\

Spin 1      &   s-wave spin group with total angular momentum \& parity: $3.0^+$    \\
Spin 2      &   s-wave spin group with total angular momentum \& parity:  $4.0^+$    \\

$\langle D \rangle_1 $ , $\langle D \rangle_2 $  & 9.0, 8.3 eV   \\
$\langle \gamma_{\gamma}^2 \rangle_1 $, $\langle \gamma_{\gamma}^2 \rangle_2 $ & 32.0, 32.0 meV \\
$\langle \gamma_{n}^2 \rangle_1$, $\langle \gamma_{n}^2 \rangle_2$ & 452.6, 332.2 meV \\
\bottomrule
\end{tabular}
\end{table}

\begin{table}[H]
\label{tabel:JEFF_par}
\caption{Prior and posterior resonance parameter for JEFF-3.3 model.
Rows in gray indicate external resonances generated using assumptions about average parameters (not informed by prior model).
}
\begin{tabular}{ccccccc}
\toprule
$E_{\lambda}$ & $\Gamma_{\gamma}$ & $\Gamma_n$ & 
$E_{\lambda}'$ & $\Gamma_{\gamma}'$ & $\Gamma_n'$ & 
% $\mathrm{spin_{ID}}$ \\
spin \\
\midrule

\rowcolor{gray!20} % Highlighted row
243.79 & 64.00 & 46.50      & 243.79 & 62.97 & 11.39    & 1.00 \\
\rowcolor{gray!20} % Highlighted row
188.81 & 64.00 & 46.50      & 188.81 & 64.08 & 78.69    & 1.00 \\
\rowcolor{gray!20} % Highlighted row
239.99 & 64.00 & 35.50      & 239.99 & 56.69 & 19.56    & 2.00 \\
\rowcolor{gray!20} % Highlighted row
192.61 & 64.00 & 35.50      & 192.61 & 65.46 & 242.95   & 2.00 \\

200.00 & 63.00 & 31.42      & 200.13 & 67.44 & 35.18    & 1.00 \\
204.67 & 65.00 & 2.754      & 204.74 & 30.02 & 2.94     & 1.00 \\
208.48 & 65.00 & 9.422      & 208.52 & 46.42 & 10.40    & 2.00 \\
215.09 & 65.00 & 48.27      & 215.10 & 44.30 & 56.03    & 2.00 \\
216.60 & 65.00 & 19.39      & 216.64 & 47.03 & 18.97    & 1.00 \\
219.81 & 65.00 & 12.88      & 219.86 & 33.30 & 13.96    & 2.00 \\
222.29 & 65.00 & 2.44       & 222.30 & 46.61 & 2.80     & 2.00 \\
225.15 & 65.00 & 20.57      & 225.20 & 54.80 & 24.05    & 2.00 \\
230.62 & 65.00 & 22.42      & 230.64 & 16.42 & 21.35    & 2.00 \\
232.26 & 65.00 & 40.17      & 232.32 & 65.03 & 43.88    & 2.00 \\
\bottomrule
\end{tabular}
\end{table}

\begin{table}[H]
\label{tabel:ENDF_par}
\caption{Prior and posterior resonance parameter for ENDF-8.0 model.
Rows in gray indicate external resonances generated using assumptions about average parameters (not informed by prior model).
}
\begin{tabular}{ccccccc}
\toprule
$E_{\lambda}$ & $\Gamma_{\gamma}$ & $\Gamma_n$ & 
$E_{\lambda}'$ & $\Gamma_{\gamma}'$ & $\Gamma_n'$ & 
% $\mathrm{spin_{ID}}$ \\
spin \\
\midrule

\rowcolor{gray!20} % Highlighted row
243.79 & 64.00 & 46.50      & 243.79 & 63.95 & 11.62    & 1.00 \\
\rowcolor{gray!20} % Highlighted row
188.81 & 64.00 & 46.50      & 188.81 & 59.15 & 44.38    & 1.00 \\
\rowcolor{gray!20} % Highlighted row
239.99 & 64.00 & 35.50      & 239.99 & 64.62 & 5.72     & 2.00 \\
\rowcolor{gray!20} % Highlighted row
192.61 & 64.00 & 35.50      & 192.61 & 42.63 & 288.30   & 2.00 \\

200.00 & 62.00 & 35.55      & 200.12 & 86.35 & 24.44    & 2.00 \\
204.70 & 56.00 & 2.97       & 204.74 & 28.29 & 2.95     & 1.00 \\
208.40 & 56.00 & 9.77       & 208.51 & 49.09 & 10.40    & 2.00 \\
215.00 & 65.00 & 54.85      & 215.12 & 44.21 & 69.83    & 1.00 \\
216.60 & 56.00 & 16.00      & 216.64 & 50.17 & 14.94    & 2.00 \\
219.70 & 56.00 & 19.08      & 219.86 & 43.10 & 17.96    & 1.00 \\
222.30 & 52.00 & 1.95       & 222.30 & 64.15 & 2.91     & 2.00 \\
225.30 & 56.00 & 25.71      & 225.20 & 65.60 & 30.21    & 1.00 \\
230.50 & 56.00 & 16.00      & 230.64 & 16.32 & 20.99    & 2.00 \\
232.30 & 56.00 & 57.77      & 232.32 & 75.07 & 43.58    & 2.00 \\

\bottomrule
\end{tabular}
\end{table}

\begin{table}[H]
\label{tabel:AutoFit_par}
\caption{Final (posterior) resonance parameters for the selected (10-resonance) model automatically with without a prior.
Rows in gray indicate external resonances generated using assumptions about average parameters.
}
\begin{tabular}{lccc}
\toprule
$E_{\lambda}'$ & $\Gamma_{\gamma}'$ & $\Gamma_n'$ & 
spin \\
\midrule

\rowcolor{gray!20} % Highlighted row
243.790 & 89.373  & 0.431   & 1.000 \\
\rowcolor{gray!20} % Highlighted row
188.810 & 24.414  & 35.148  & 1.000 \\
\rowcolor{gray!20} % Highlighted row
239.990 & 455.267 & 0.220   & 2.000 \\
\rowcolor{gray!20} % Highlighted row
192.610 & 43.614  & 267.406 & 2.000 \\

200.145 & 51.343  & 38.041  & 1.000 \\
204.742 & 44.002  & 3.064   & 1.000 \\
215.133 & 48.962  & 67.342  & 1.000 \\
216.622 & 57.798  & 19.577  & 1.000 \\
222.310 & 56.852  & 3.765   & 1.000 \\
225.210 & 44.167  & 32.895  & 1.000 \\
232.336 & 51.396  & 58.310  & 1.000 \\
208.514 & 35.065  & 10.719  & 2.000 \\
219.867 & 57.101  & 13.518  & 2.000 \\
230.629 & 29.061  & 19.990  & 2.000 \\
\bottomrule
\end{tabular}
\end{table}

% \label{sec:sample:appendix}
\newpage
\section*{Acknowledgements}
% NSSC acknowledgements
This material is based upon work supported by the Department of Energy National Nuclear Security Administration through the Nuclear Science and Security Consortium under Award Number(s) DE-NA0003996.
\vskip 10pt 
\noindent
% Jesse's acknowledgements
This work was supported by the Nuclear Criticality Safety Program, funded and managed by the National Nuclear Security Administration for the U.S. Department of Energy. Additionally, work at Brookhaven National Laboratory was sponsored by the Office of Nuclear Physics, Office of Science of the U.S. Department of Energy under Contract No. DE-SC0012704 with Brookhaven Science Associates, LLC. This project was supported in part by the Brookhaven National Laboratory (BNL), National Nuclear Data Center under the BNL Supplemental Undergraduate Research Program (SURP) and by the U.S. Department of Energy, Office of Science, Office of Workforce Development for Teachers and Scientists (WDTS) under the Science Undergraduate Laboratory Internships Program (SULI).
\vskip 10pt 
\noindent
% BNL acknowledgements
This work was supported by the Nuclear Criticality Safety Program, funded and managed by
the National Nuclear Security Administration for the Department of Energy
\vskip 10pt 
\noindent
This report was prepared as an account of work sponsored by an agency of the United States Government. Neither the United States Government nor any agency thereof, nor any of their employees, makes any warranty, express or implied, or assumes any legal liability or responsibility for the accuracy, completeness, or usefulness of any information, apparatus, product, or process disclosed, or represents that its use would not infringe privately owned rights. Reference herein to any specific commercial product, process, or service by trade name, trademark, manufacturer, or otherwise does not necessarily constitute or imply its endorsement, recommendation, or favoring by the United States Government or any agency thereof. The views and opinions of authors expressed herein do not necessarily state or reflect those of the United States Government or any agency thereof.
\vskip 10pt 
\noindent
The authors also gratefully acknowledge Dr. Vitaly Ganusov (University of Tennessee) for valuable discussions and insightful feedback during the course of this research regarding model selection approaches.
\newpage
 \bibliographystyle{elsarticle-num} 
 \bibliography{cas-refs}





% \end{thebibliography}
\end{document}